\documentclass[preprint,pra,aps]{revtex4}

\usepackage{graphicx}
\usepackage{dcolumn}   
\usepackage{bm,braket}        
\usepackage{amssymb}   
\usepackage{amsmath}
\usepackage{url,color}
\usepackage{circuitikz}
\usepackage{qcircuit}
\usepackage{layout}
\usepackage{epsfig}
\usepackage{graphicx}
\usepackage{epstopdf}
\usepackage{booktabs}
\usepackage{float,CJK}

\def\be{\begin{equation}}
\def\ee{\end{equation}}
\def\ba{\begin{eqnarray}}
\def\ea{\end{eqnarray}}
\def\sdg{Schr\"odinger~}

\begin{document}
\title{Classical Computer, Quantum Computer, and the G\"odel's theorem}
\author{Biao Wu}
\email{wubiao@pku.edu.cn}
\affiliation{International Center for Quantum Materials, School of Physics, Peking University,  Beijing 100871, China}
\affiliation{Wilczek Quantum Center, School of Physics and Astronomy, Shanghai Jiao Tong University, Shanghai 200240, China}
\affiliation{Collaborative Innovation Center of Quantum Matter, Beijing 100871,  China}
\date{\today}

\begin{abstract}
I show that the cloneability of information is the key difference between classical computer and quantum computer. 
As information stored and processed by neurons is cloneable, brain (human or non-human) is a classical computer. 
Penrose argued with the G\"odel theorem that human brain is not classical. I demonstrate with an example why 
his argument is flawed. At the end, I discuss how to go beyond quantum computer.   
\end{abstract}

\maketitle

Galileo wrote in 1623, ``{\it Philosophy is written in this grand book - I mean the universe - which stands continually open to our gaze, but it cannot be understood unless one first learns to comprehend the language in which it is written. It is written in the language of mathematics, and its characters are triangles, circles, and other geometric figures, without which it is humanly impossible to understand a single word of it; without these, one is wandering about in a dark labyrinth.  }" ~\cite{assayer} This was a remarkably bold and visionary statement at the time when there were 
only a few accurate mathematical descriptions of physical phenomena, such as  free fall  and pendulum. Currently
there is no doubt  that the universe is written in the language of mathematics.  A physical process is literally a demonstration 
of  a mathematical solution.  As computation is a  process of finding mathematical solutions, a physical process can be regarded as a  
computation of a mathematical equation.  Or simply, physics is computational. How about the reverse,  is computation physical? 
The answer had been `No'  until the discovery of quantum computer~\cite{Manin,Feynman,Deutsch1,Deutsch2}. 
It is now widely accepted that computation is physical. Different models of computation correspond to different physical laws. 
Classical physics underlies Turing machine and other models of classical computer; quantum physics is behind quantum computer.  

However, it is not obvious at all why classical computer is classical and quantum computer is quantum. 
Neither is there the Newton's equations of motion
explicitly in classical computer nor is there the \sdg equation explicitly in quantum computer. 
There is even no Planck constant $\hbar$ in theoretical models of quantum computer.  It turns out that 
being classical and being quantum are   concepts more fundamental than equations of motion. 
The nature of a computer  is determined by the cloneability of the information processed by the computer. 
If the information is cloneable,  the computer is classical; if it is uncloneable, the computer is quantum. 
As the information processed in neurons are cloneable, brains (human or non-human) are classical computers. 
Penrose argued that the G\"odel theorem implies that human brains (not all brain) are not classical~\cite{Penrose}. I shall show
that this is due to his misunderstanding of the theorem. The G\"odel theorem in fact implies that there are innumerable 
mathematical statements so that human mathematicians, who have only finite physical resources, cannot prove all
of them true or false. An example is constructed to demonstrate it. In the end, I shall give an example on how to 
construct theoretical models of computing that are more powerful than quantum computer~\cite{HWW}. To do this, 
the information processed by the computer has to be different from both classical computer and quantum computer.

\section{Classical computer and quantum computer}
There are many different models of classical computer, such as the famed Turing machine and the conventional digital 
computer. They are all equivalent in terms of theoretical computing power~\cite{nielson2000quantum}.  
I shall use the reversible classical computer to illustrate the 
essential features of  classical computer. The primary reason is that quantum computer 
can be viewed as a straightforward extension or quantization of reversible classical computer. 

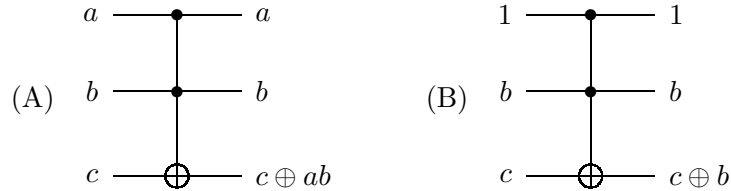
\begin{figure}[h]
(A)~~~~~~\begin{minipage}{4cm}
\Qcircuit @C=1.8em @R=2.4em {
		\lstick{a}	& \ctrl{1}   &\rstick{a} \qw\\	 
		\lstick{b}   &  \ctrl{1}  &\rstick{b} \qw\\
		\lstick{c}	 & \targ       &\rstick{c\oplus ab} \qw}
\end{minipage}
\hspace{2.2cm}
(B)~~~~~~\begin{minipage}{4cm}
\Qcircuit @C=1.8em @R=2.4em {
		\lstick{1}	    & \ctrl{1} & \rstick{1} \qw\\	 
		\lstick{b} 	    &  \ctrl{1} & \rstick{b}\qw \\
		\lstick{c}	    & \targ & \rstick{c\oplus b}\qw}
\end{minipage}
	\caption{
	(A) the Toffoli gate; (B) CNOT gate constructed from the Tofolli gate. The inputs $a,b,c$ are Boolean
	variables.}
	\label{Tofolli}
\end{figure}

A reversible classical computer has  one universal logic gate,  the Tofolli gate, along with an array of bits. 
Each bit has only two values, 0 and 1. 
The Tofolli gate, as shown in  Fig.\ref{Tofolli}B, has two control bits and one target bit. Only when both the control bits are 1, 
the target bit switches between 0 and 1~\cite{nielson2000quantum}.

On top of the reversible classical computer, if we add two more logic gates and replace  bits with qubits, we have the standard
circuit model of quantum computer. The two additional gates are the Hadamard gate $H$ and  $\pi/8$ gate $T$~\cite{nielson2000quantum},
\be
H=\frac{1}{\sqrt{2}}\begin{pmatrix}
1&1\\ 1&-1
\end{pmatrix}~\,,~~~~~T=\frac{1}{\sqrt{2}}\begin{pmatrix}
1&0\\ 0& e^{i\pi/4}
\end{pmatrix}~\,.
\ee
The usual  universal quantum gate, CNOT gate, can be easily constructed from the Tofolli gate as shown Fig.\ref{Tofolli}(B). The key difference
between qubit and bit is that a qubit can be in a superposition state
\be
\ket{\psi}=\alpha\ket{0}+\beta\ket{1}\,.
\ee
The states of a bit correspond to two special cases, $\alpha=1,\beta=0$ and $\alpha=0,\beta=1$. 

The Hadamard gate is the key in quantum computer. It is the only universal gate that is capable of generating a superposition state
out of a non-superposition state, for example, 
\be
H\ket{0}=\frac{1}{\sqrt{2}}(\ket{0}+\ket{1})\,.
\ee
Only with Hadamard gates will we be able to generate more superposition terms, change 
the superposition amplitudes, and facilitate interference  in the course of computation. This shows that quantum computer 
can be viewed as a quantization of reversible classical computer. The quantization is achieved primarily 
with the Hadamard gate and additionally with $\pi/8$ gate $T$, but remarkably without Planck constant.  
A classical computer with $n$ bits have exactly $2^n$ states, 
\be
\ket{J}=\ket{J_1,J_2,\cdots,J_k,\cdots,J_n}
\ee
where $J_k$ is the $k$th binary digit of $J$ ($J=0, 1, 2,\cdots, 2^{n}-1$).  In contrast, a quantum computer with $n$ qubits has infinite  
states that live in a Hilbert space spanned by $\ket{J}$'s. In general, a quantum computer is in a superposition state as
\be
\ket{\Psi}=\sum_{J=0}^{2^n-1} a_J\ket{J}\,. 
\label{super}
\ee

\section{Cloneable states and uncloneable states}
\label{sec:clone}
The reversible classical computer is clearly not quantum as it does not allow generation of superposition.  
But why is it classical? Why is Turing machine classical?  
The conventional digital computer  is also regarded 
classical even it is irreversible.  What do these models of classical computer have in common with classical mechanics?  
Similarly, why is the model of quantum computer described above quantal? There is even no Planck constant in the model. 
 
To answer these questions,  let us review dynamical systems in general. A dynamical system 
has a configuration space and an evolution rule (or a set of evolution rules). Its state is a point in the configuration space. 
The system can start at any point in the configuration space. The evolution rule dictates it 
 to evolve deterministically  from one state to another. Classical mechanics is a dynamical system. The configuration space
is the phase space and the evolution rule is the Newton's equations of motion. Quantum mechanics is a dynamical system. 
The configuration space is the Hilbert space and the evolution rule is the \sdg equation. 
Mathematically, classical mechanics and quantum mechanics, are essentially the same type of dynamical systems. 
Koopman and von Neumann formulated  classical mechanics mathematically as quantum systems~\cite{Koopman,Neumann}. 
And Heslot converted  the \sdg equation mathematically into a set  of Newton's equations of motion~\cite{Heslot}. 

This shows that the equations of motion are not what makes  classical mechanics classical and quantum mechanics quantal. 
The key difference is how their states, the points in their configuration spaces, are related to physical reality.  
The coordinates of a point in the phase space are momenta and positions, which can be directly measured and cloned. 
In contrast,  a point in a Hilbert space, usually called wave function, is not directly observable and can not be cloned in general. 
To relate a wave function  to physical reality, one has to first select a complete set of orthonormal basis, which are usually the eigenfunctions of 
a physical quantity such as momentum, position or angular momentum. This selection of basis is outside of the configuration
space and the equations of motion, and is done independently.  When the wave function is expanded in the basis, 
the expansion coefficients ( or the coordinates)  are not measurable,  and thus not cloneable.  Upon measurement, 
the  observed is the basis vector and its corresponding eigenvalue, not the coefficients ( or the coordinates). 
To distinguish this crucial difference, we call classical state ( a point in phase space ) {\it cloneable} and 
quantum state (a vector in  Hilbert space) 
{\it uncloneable}.  It is this cloneability of their states that separates a classical system from a quantum system, 
not its equations of motion. This cloneability is related to the well-known no-cloning theorem for quantum states\cite{Park1970,Wootters1982}.
According to this theorem,  all the  states $\ket{J}$ of  bits can be cloned but not their superpositions, such as the states
in Eq.(\ref{super}).

The states of a classical particle are cloneable. Its position and momentum can be measured and recorded directly, 
ultimately cloned faithfully with negligible errors. 
The states $\ket{J}$ of bits are cloneable not only theoretically but also in practice: the two states, 0 or 1,  of a bit, often represented by voltages, 
can be measured and observed directly, and cloned faithfully. It is this common feature with a classical particle that makes
bits classical.  This is the key reason  that  any computer with bits, such as the Turing machine,  the conventional digital computer, 
and the reversible computer  are classical.  In contrast, the states of qubits are vectors in  the Hilbert space
just as the wave functions in quantum mechanics.   Upon measurement, only partial information 
of these  states can be recorded; they can not be cloned faithfully with negligible errors. As a result, computers 
with qubits are quantum computer. It is now clear. Classical computer is classical because its states (or information) are cloneable. 
Quantum computer is quantal is because its states (or information) are uncloneable. One can force
the quantum states  be cloneable. This is exactly what happens  when we simulate a quantum system on a classical computer.  
The price to pay  is that the memory size of a classical computer would have to grow exponentially 
with the size of the quantum system.  It is precisely what motivated the conception of quantum computer~\cite{Manin,Feynman}.  
Note that one can also repeat measurements on a given quantum state and ultimately 
clone it faithfully.  Unfortunately, the number of measurements needed grows exponentially with  the system size~\cite{HYHu}. 

\section{Brain and the G\"odel's theorem}
Brain (human or non-human) is clearly a classical computer because  the states of each neuron are cloneable. 
A neuron state is marked by electrical pulses (usually called action potential)  passing through the axon, 
opening of ion channels, sodium and potassium concentrations, activation of chemical or electric signals at synapses. 
These physical and chemical states can be measured and recorded directly, and cloned faithfully at least in principle. 
Technically, it is challenging to clone faithfully the functions of a single neuron, not to mention a network of neurons (brain) . 
But it is an issue of technology, not an issue of theoretical principles. Technically, we are still far away building a machine
that moves in a jungle as adeptly as monkeys. Nevertheless, the monkey's brain, muscle, and other body parts function 
classically. 

That human brain is a classical computer is also evident in the process of a mathematician proving a theorem. 
When a mathematician proves a theorem,  axioms or assumptions are inputs, mathematical 
inductions are logical gates, known theorems or lemmas are subroutines. A completed proof is essentially a classical algorithm
because all the symbols and induction rules can be written down without any ambiguity, 
and be cloned faithfully from mathematician to mathematician. 
In other words, they are cloneable just as  the states of bits. 

However, Penrose argued the contrary that human brain is not a classical computer 
and may be of quantum nature~\cite{Penrose}. He believes that this is implied in the G\"odel's incomplete theorem, which 
states that there are statements in any consistent formal system $F$ which can neither be proved nor disproved in $F$. 
Let me first summarize Penrose's arguments or proof~\cite{Penrose}. He listed all the computations $C_q(n)$ that take a natural number $n$ 
as its input.  $C_q(n)$ is the $q$th algorithm that computes on the input $n$. $A(q,n)$ of two inputs $q$ and $n$ 
is an algorithm that intends to ascertain whether $C_q(n)$ will stop: if $A(q,n)$ stops, $C_q(n)$ does not stop. 
With Cantor's diagonalization argument, Penrose arrives at the statement (M): if $C_k(k)$ stops, $C_k(k)$ does not stop.
This leads him to  conclude that 
\begin{quote}
Human mathematicians are not using a knowably sound algorithm in order to ascertain mathematical truths. 
\end{quote}
This means that human brain is not a classical computer. For me,  (M) is simply a self-contradictory statement and means
that at least one of the assumptions made at the start of Penrose's proof is wrong. In fact, Penrose implicitly assumed that
all the algorithms $C_q(n)$ are  countable and therefore can be listed. This assumption is certainly wrong. In  Cantor's proof that 
the real numbers are innumerable, he first assumed that  the real numbers are countable and can be listed. With 
the  diagonalization argument, he found a real number outside of the list. A contradiction so that the assumption 
that the real numbers are countable is wrong.  Cantor's proof also implies that irrational numbers can not be exhausted 
by a finite number of different operations with finite symbols. In contrast,  all rational numbers can be obtained by one operation (division) 
with 10 digits (0,1,2,3,4,5,6,7,8,9).

Similarly,  it is wrong to assume that algorithms are countable. 
Here is an example that demonstrates  that there are innumerable algorithms.  
A number is said embedded in $\pi$ if its decimal digits can be found in $\pi$ in the sequential order.
For example, 1.23456 is embedded in the first 50 decimal digits of $\pi$, 
\be
3.\underline{\bm 1}4159\underline{\bm 2}65\underline{\bm 3}5 89793238\underline{\bm 4}6 2643383279 \underline{\bm 5}028841971 \underline{\bm 6}939937510...
\ee
$R^{(n)}(x)$ is an operation that repeats $n$ times each decimal digit of $x$. 
For example, $R^{(1)}(1.23456)=1.23456$ and $R^{(3)}(1.23456)=111.222333444555666$. For every real number $x$, 
we construct a list of  
mathematical statements $E^{(n)}(x)$: $R^{(n)}(x)$ is embedded in $\pi$. The statement $E^{(1)}(\sqrt{2})$ is probably true. 
Its first twelve decimal digits  are  embedded in the first 100 digits of $\pi$. 
\ba
&& 3.\underline{\bm 1}\underline{\bm 4}\underline{\bm 1}5926535 89793238\underline{\bm 4}6 \underline{\bm 2}643383279502884
\underline{\bm 1}97169\underline{\bm 3}9937\underline{\bm 5}10\nonumber\\
 &&5820974944 59230781\underline{\bm 6}4 06\underline{\bm 2}8620899 86280\underline{\bm 3}4825 34211\underline{\bm 7}0679 ...
\ea
Its thirteenth digit is $3$ and will be found in the rest digits of $\pi$, similarly its fourteenth digit $0$, and so on.  
The statement $E^{(2)}(\sqrt{2})$ is no longer as obvious. Here is the situation for the fist 200 digits of $\pi$. 
\ba
&&3.1415926535 8979323846 2643383279 5028841971 6939937510 \nonumber\\
&&5820974944 5923078164 0628620899 8628034825 342\underline{\bm {11}}70679 \nonumber\\
&&8214808651 3282306647 
0938\underline{\bm{44}}6095 5058223172 5359408128 \nonumber\\
&&48\underline{\bm{11}}174502 8410270193 8521105559 6\underline{\bm{44}}6\underline{\bm{22}}9489 5493038196 ...
\ea
But still it is hard to believe that the next two digits $11$ never appear in the rest digits of $\pi$,  similarly $33$, and so on. 
$E^{(2)}(\sqrt{2})$ is probably true. As $n$ increases, we become less confident that the statement $E^{(n)}(\sqrt{2})$ is true. 
Consider $n=6$.  The first string of six or more consecutive identical digits is six 9's starting at the 762nd decimal place of $\pi$, 
the next is again six 9's starting at position 193034, and the next is six 8's starting at position 222,299~\cite{pai}. 

To verify whether $E^{(n)}(x)$'s are true or not, one can set up algorithms that simply compute every decimal digits of $x$ and $\pi$ 
and check them one by one. This method can show that some statements, such as $E^{(6)}(9)$,  are true. When $x$ is irrational,  
this method fails. It can not even show with 100\% certainty whether  $E^{(1)}(\sqrt{2})$ is true because mathematicians (including machines) 
have only finite resources and they can not compute infinite number of digits. 
However, sometimes one does not need to compute out all the digits of an irrational number to verify  a statement true or false.  
For example, one can show that $\sqrt{2}$ is irrational and $E^{(1)}(\pi)$ is true without computing out all the digits. 
So, it is possible that mathematicians can figure out a `clever' way or algorithm to verify that $E^{(1)}(\sqrt{2})$ is true. However, it is hard
to imagine that  `clever' algorithms can be found for all irrational numbers and any $n$. 

As the real numbers are innumerable, we have innumerable  mathematical statements $E^{(n)}(x)$. 
Each statement $E^{(n)}(x)$ is clearly also an algorithm with two inputs $n$ and $x$, and 
can be identified with $C_q(n)$ via $q=x$. This example shows that
Penrose's assumption that $C_q(n)$'s are countable is wrong. 

We now face two realities: (1) there are  innumerable number of mathematical statements; (2) human mathematicians
are only capable of using finite number of symbols,  finite sets of  induction rules,
and doing finite steps of deductions ( similar to a finite number of logical gates in an algorithm). 
The latter means that the mathematical statements that can be verified by human mathematicians are countable and can be listed. 
Therefore, 
it is clear that mathematicians can not prove all the mathematical statements, which are innumerable,  true or false. 
This is the essence of the G\"odel's  theorem. Therefore, the G\"odel's  theorem has nothing to do with
whether human brain is classical or not. The theorem just shows with mathematical rigor  a common sense:  
it is impossible for any thinking machine (including human brain) with finite resources to solve innumerable problems.  

The G\"odel's  theorem is often explained with self-referencing examples, for example, this statement is not true. 
This self-referencing is made rigorous in its proof with the  diagonalization argument. 
It gives the theorem a mystique appearance that obscures its essence and true implication: innumerable problems 
can not be solved with finite physical resources. 

There are obvious differences between brain and the conventional digital computer. It is a mundane task for 
many animals, including human, to walk or run on difficult terrains; it is declared a major achievement
whenever  robots's walking abilities have been improved in one or another aspect. It takes a long time for a human to add 
two large numbers; it takes almost no time for a digital computer to do it. 
This does not imply  that  human brain may be a quantum computer. It is simply that we do not understand
the networking of neurons inside human brain. Just as we do not need to know every technical details of a digital computer 
to know for sure that it is classical, we can  be absolutely sure that human brain is classical before we understand it
completely. 

Galileo said that physics is written in the language of mathematics; G\"odel proved that mathematics is limited by physics. 
After all, human is physical. All human products, including mathematical theorems, are limited by the physical laws. 
What we know and will know is negligible  relative to 
what we will never know.  This is in a sense good as we will never run out of truths to discover.

\section{Beyond quantum computing}
Is it possible to go beyond quantum computer?  According the analysis in Section \ref{sec:clone}, 
it is impossible in practice since we do not have any experimentally tested theory that goes beyond quantum mechanics. 
Theoretically, it is certainly possible. We only need a dynamical system whose states in its configuration space
are related to physical reality differently from both classical mechanics and quantum mechanics. 

One of such systems was discussed long time ago by Pauli~\cite{Pauli1943} who was inspired by Dirac~\cite{Dirac1942}.
It was considered independently 
many years later as a generalization of Bogoliubov-de Gennes equation~\cite{zhang2018lorentz}. It was named 
Lorentz quantum mechanics because its dynamical evolution is generalized Lorentz transformation. 
The  key features of the Lorentz mechanics are:
(1) its states are vectors in a complex linear space with  an indefinite metric; (2) its dynamics is generalized 
Lorentz transformation; (3) there are two sets of states, one with positive inner products and the other with
negative products.  The third feature is crucial: the positive states are regarded as physical 
and can be measured; the negative one are regarded unphysical. This means that the states in the configuration space, 
an inner product space with indefinite metric, are not only uncloneable and half of them are also unphysical. 
This is different from both classical states and quantum states. 

We have proposed a new theoretical model of computing based on this Lorentz quantum mechanics~\cite{HWW}. 
We find that the search algorithm designed for this new computer model is exponentially faster than the Grover's 
algorithm~\cite{Grover}. It shows explicitly that one can go beyond conventional quantum computer. 
If  in a new theory that unifies quantum theory and gravity its states  are different from both
classical states and quantum states, we would also expect to have a more powerful model of computing. 

There are two computer models that were shown to be more powerful than  quantum 
computer~\cite{abrams1998nonlinear,bacon2004quantum}. Both add nonlinearity to quantum mechanics
and claim that nonlinearity is behind the enhancement of computing power. It is suspicious argument. 
Classical mechanics is inherently nonlinear and no one has been able to exploit its nonlinearity to enhance computing power. 
It is no surprise that both models were critically questioned from the theoretical point of view\cite{Bennett2009}. \\

\acknowledgments
I thank Zhenduo Wang, Yu Shi, and Zhuoquan Lin for helpful discussion. 
This work is supported by the The National Key R\&D Program of China (Grants No.~2017YFA0303302, No.~2018YFA0305602), National Natural Science Foundation of China (Grant No. 11921005), and 
Shanghai Municipal Science and Technology Major Project (Grant No.2019SHZDZX01).

\end{document}